\documentclass[useAMS,usenatbib,usegraphicx]{pasj00}
\usepackage{lscape}
\usepackage{color}
\usepackage{natbib, aas_macros}
\Received{$\langle$reception date$\rangle$}
\Accepted{$\langle$acception date$\rangle$}
\Published{$\langle$publication date$\rangle$}
\SetRunningHead{Astronomical Society of Japan}{Radio Relics Tracing the Projected Mass Distribution in CIZA J2242.8+5301}

\title{Radio Relics Tracing the Projected Mass Distribution in CIZA J2242.8+5301
\thanks{Based on data collected at Subaru Telescope, which is operated by the National Astronomical Observatory of Japan.}}

 \author{%
N. \textsc{Okabe}, \altaffilmark{1,2,3}
H. \textsc{Akamatsu},\altaffilmark{4}
J. \textsc{Kakuwa}, \altaffilmark{1}
Y. \textsc{Fujita}, \altaffilmark{5}
Y. -Y. \textsc{Zhang}, \altaffilmark{6}
M. \textsc{Tanaka}, \altaffilmark{7}
and
K. \textsc{Umetsu}, \altaffilmark{8}
}
\altaffiltext{1}{Department of Physical Science, Hiroshima University,
1-3-1 Kagamiyama, Higashi-Hiroshima, Hiroshima 739-8526, Japan}
\altaffiltext{2}{Hiroshima Astrophysical Science Center, Hiroshima University, Higashi-Hiroshima,
Kagamiyama 1-3-1, 739-8526, Japan}
\altaffiltext{3}{Kavli Institute for the Physics and Mathematics of the Universe (WPI), Todai Institutes for Advanced Study,University of Tokyo, 5-1-5 Kashiwanoha, Kashiwa, Chiba 277-8583, Japan}
 \email{okabe@hiroshima-u.ac.jp}
\altaffiltext{4}{SRON Netherlands Institute for Space Research, Sorbonnelaan 2, 3584 CA Utrecht, The Netherlands}
\altaffiltext{5}{Institute of Space and Astronautical Science, Japan Aerospace Exploration Agency,3-1-1 Yoshinodai, Chuo-ku, Sagamihara, Kanagawa 229-8510, Japan}
\altaffiltext{6}{Argelander-Institut f\"ur Astronomie, Universit\"at Bonn, Auf dem H\"ugel 71, 53121 Bonn, Germany}
\altaffiltext{7}{National Astronomical Observatory of Japan, Mitaka, Tokyo 181-8588, Japan}
\altaffiltext{8}{Institute of Astronomy and Astrophysics, Academia
Sinica, P.O. Box 23-141, Taipei 10617, Taiwan}

\newcommand{\simgt}{\lower.5ex\hbox{$\; \buildrel > \over \sim \;$}}
\newcommand{\simlt}{\lower.5ex\hbox{$\; \buildrel < \over \sim \;$}}

\def\hkpc{\mathrel{h^{-1}{\rm kpc}}}

\def\hMsol{\mathrel{h^{-1}M_\odot}}
\def\h70Msol{\mathrel{h_{70}^{-1}M_\odot}}
\begin{document}

\date{\today}


\maketitle

\label{firstpage}

\begin{abstract}
We present a weak-lensing analysis for the merging galaxy cluster, CIZA
 J2242.8+5301, hosting double radio relics, using three-band Subaru/Suprime-Cam
 imaging ($Br'z'$). 
Since the lifetime of dark matter halos colliding into clusters is longer than that of X-ray
 emitting gas halos, weak-lensing analysis is a powerful method to constrain a merger
 dynamics. Two-dimensional shear fitting using a clean background catalog 
suggests that the cluster undergoes a merger with a mass ratio of about 2:1. 
The main halo is located around the gas core in the southern
 region, while no concentrated gas core is associated with the northern sub halo.
We find that the projected cluster mass distribution resulting from an
 unequal-mass merger is in excellent agreement with the curved shapes of
 the two radio relics and the overall X-ray morphology except for the
 lack of the northern gas core.
The lack of a prominent radio halo enables us to constrain an upper limit of the
 fractional energy of magneto-hydrodynamics 
 turbulence of $(\delta
 B/B)^2<\mathcal{O}(10^{-6})$ at a resonant wavenumber, 
by a balance between the acceleration time and the time after the core
 passage or the cooling time, with an assumption of resonant acceleration 
by second-order Fermi process.
\end{abstract}

\makeatletter

\section{Introduction}

The acceleration mechanism generating relativistic particles 
in the intracluster medium (ICM) along with
its associated non-thermal phenomena is one of the long-standing
unresolved problems in studies of cluster physics. 
The presence of relativistic electrons and magnetic fields is recognized by
diffuse, synchrotron radio emission (on Mpc scales),
which is not associated with compact sources such as AGN, but with cluster
merging \citep[e.g.][]{Feretti12,Brunetti14}. 
The diffuse radio emission in galaxy clusters is classified according 
to their morphology into two types, radio halos and relics.
Radio halos, to first order, follow the X-ray brightness distribution in the central region, 
sometimes along with a curved spectrum \citep[e.g.][]{Brunetti08}. 
Radio relics display a filamentary structure in the peripheral region.
Some relics have a harder spectrum than that of radio halos
\citep[e.g.][]{Brunetti08, vanWeeren10}.
The fundamental physics to form radio halos and relics and 
give rise to different morphological types are one of the most outstanding problems. 
Furthermore, the acceleration mechanism is still unclear from the theoretical side. 
In theory, two types of acceleration mechanism have been proposed, the first
\citep[e.g.][]{Bell78,Blandford78,Drury83}
and second order Fermi acceleration \citep[e.g.][]{Schlickeiser87,Ohno02,Fujita03,Brunetti04}. 
It is unresolved through what process relativistic particles are
accelerated/re-accelerated in merging clusters.
Since the merger dynamics is controlled by dark matter, 
it is of vital importance for understanding non-thermal phenomena 
to measure the cluster mass distribution for merging clusters.
\citep[e.g.][]{Okabe08,Okabe10b,Okabe11,Medezinski15}.

Recent radio observation \citep{vanWeeren10} discovered spectacular,
double-giant radio relics in CIZA J2242.8+5301. 
The northern relic is larger than the southern one, which
suggests non-equal-mass merger \citep{vanWeeren11}.
The northern giant relic shows 
a spatial distribution of aging relativistic electrons, 
indicating a presence of strong magnetic field ($5-7\mu$G). 
The magnetic field at the radio relic is stronger than equipartition magnetic fields in radio
halos ($\sim0.1-1\mu$G), but consistent with a lower limit of the
magnetic field ($>3\mu$G) in other relics by comparing the inverse Compton X-ray
emission with the measured radio synchrotron emission \citep[e.g.][]{Nakazawa09,Finoguenov10}.
Assuming a diffusive shock acceleration  \citep[DSA;][]{Drury83}, the Mach
number of shock velocity is expected to be ${\mathcal
M}=4.6^{+1.3}_{-0.9}$. 
{\it Suzaku} X-ray observation \citep{Akamatsu13}
has discovered a temperature jump across the northern relic and estimated
that the Mach number is somewhat lower than expected by radio observation.   
A discrepancy between the X-ray-estimated and radio-estimated Mach
numbers makes it puzzling to understand a particle
acceleration process.
On the other hand, \cite{Stroe14b} have pointed out that 
an intrinsic beam-image with of the radio telescope affects the
estimation of the radio spectral index and revised ${\mathcal
M}=2.90^{+0.10}_{-0.13}$ which is close to X-ray measurements.
Therefore, the study on a comparison of Mach numbers is still controversial. 
No prominent radio halo in the central region is 
found by multi-bands radio analysis \citep{vanWeeren10,Stroe13}.
Recent study of weak-lensing analysis \citep{Jee15} has shown a bimodal
structure of the dark matter and member galaxies and concluded an
equal-mass merger.

We here report weak-lensing analysis using data from
Subaru/Suprime-Cam. 
The paper is organized as follows: Section 2
presents the weak-lensing analysis of Subaru/Suprime-Cam. 
We present distribution of member galaxies and mass
measurements in Sections 3 and 4, respectively. 
In Section 5, we compare the projected mass distribution with the X-ray
emitting gas, radio relics and member galaxy distribution. 
The results are discussed in Section 6 and summarized in Section 7. 
We use $\Omega_{m,0}=0.3$, $\Omega_{\Lambda}=0.7$ and $H_0=100h~{\rm
 km~s^{-1}}Mpc^{-1}$ ($H_0=70h_{70}~{\rm km~s^{-1}Mpc^{-1}}$).

\section{Data Analysis} \label{sec:data}

We took three-band imaging data ($B,r',z'$) for CIZAJ2242.8+5301 ($z_l=0.192$)
with the Suprime-cam \citep{Miyazaki02} at the Subaru 8.2-m telescope. 
Here, the subscript $l$ denotes the lensing (cluster) redshift.
The exposure times for the $B$- $r'$- and $z'$- bands are $24$, $36$ and
$12$ mins, respectively.
Those imaging are taken through the Service program (P.I.:N. Okabe).
The $r'$- image, of which seeing is $0.7$ arcsec, 
is used to measure ellipticities of galaxies.
The other bands are combined with the $r'$-band data to minimize a contamination of unlensed member
galaxies in background source catalog. 

We use the standard pipeline reduction software for the Suprime-Cam, 
SDFRED \citep{Yagi02,Ouchi04} modified for the new CCDs.
The data reduction follows \cite{Okabe13,Okabe14a}. 
The zero point of magnitude is corrected for the Galactic extinction
\citep{Schlafly11} using a single function over the whole field and calibrated 
by comparing magnitudes of stellar objects with those of standard stars

We measure ellipticity of each galaxy in terms of weighting the surface
brightness, following the KSB method \citep{KSB} with some modifications \citep{Okabe14a}.
The details are described in \cite{Okabe14a}. 
The photometric redshift of each galaxy is estimated as the ensemble
average of the COSMOS photometric redshift \citep{Ilbert13} of neighboring
COSMOS galaxies in the three-magnitudes plane. 
We adopt 100 COSMOS galaxies for the estimation and confirm that 
the change by changing the sampling number is negligible. 
The galaxies for shape measurements are selected by
$r_h>{\bar r}_h^*+\delta{r_h}^*$, $r_g>{\bar r}_g^*+\delta{r}_g^*$ and $22<r'$, 
where $r_h$ and $r_g$ are the half-light radius and the Gaussian radius,
respectively. The asterisk denotes the stellar objects. Since the cluster
is located at the Galactic plane, the number density of background galaxies is
small $9~{\rm arcmin^{-2}}$ because of a considerable number of
contaminated stars, compared to those in previous studies 
\citep[e.g.][]{Okabe08,Okabe10b,Okabe13,Umetsu14}.

\begin{figure*}
\includegraphics[width=0.46\hsize]{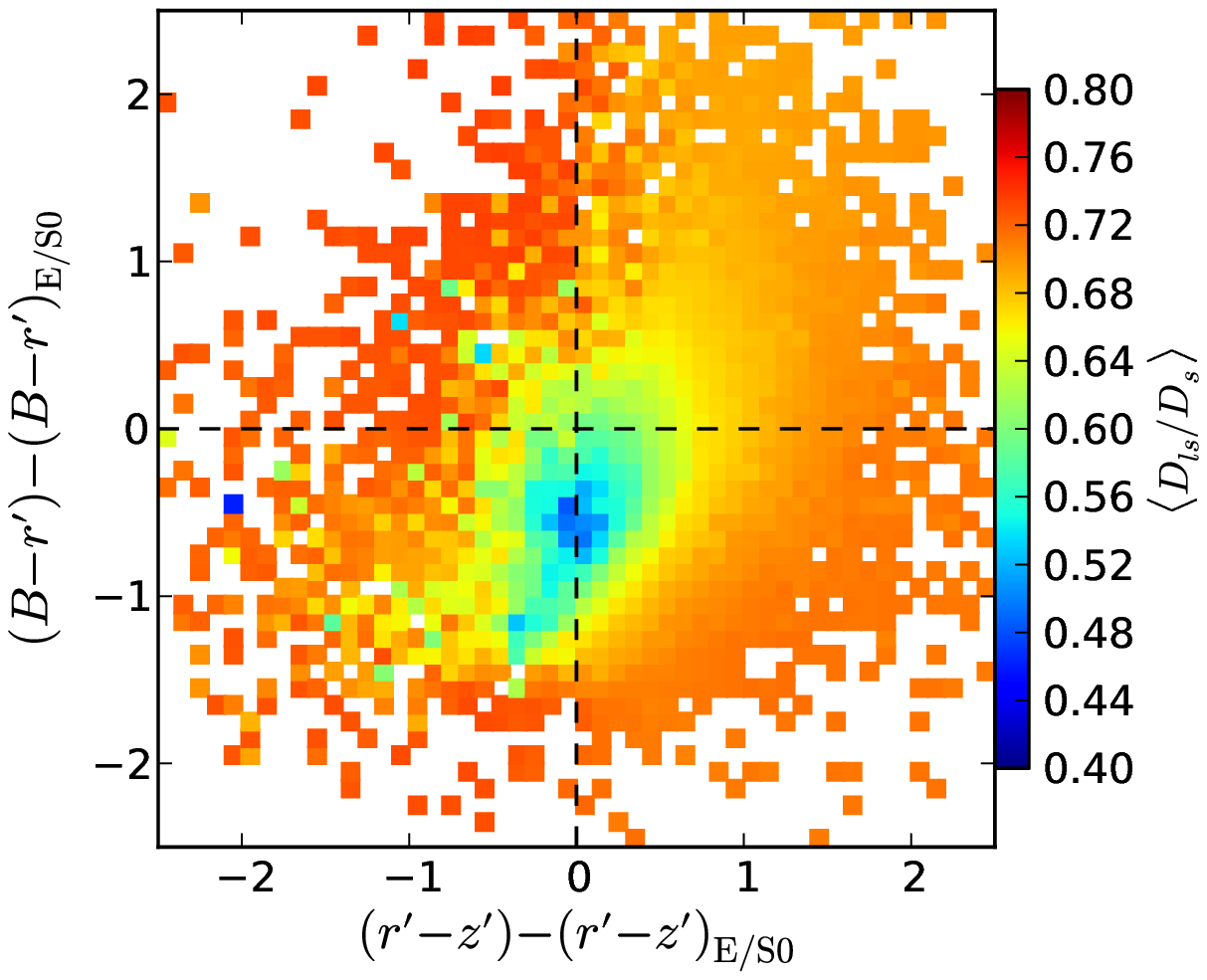}
\includegraphics[width=0.45\hsize]{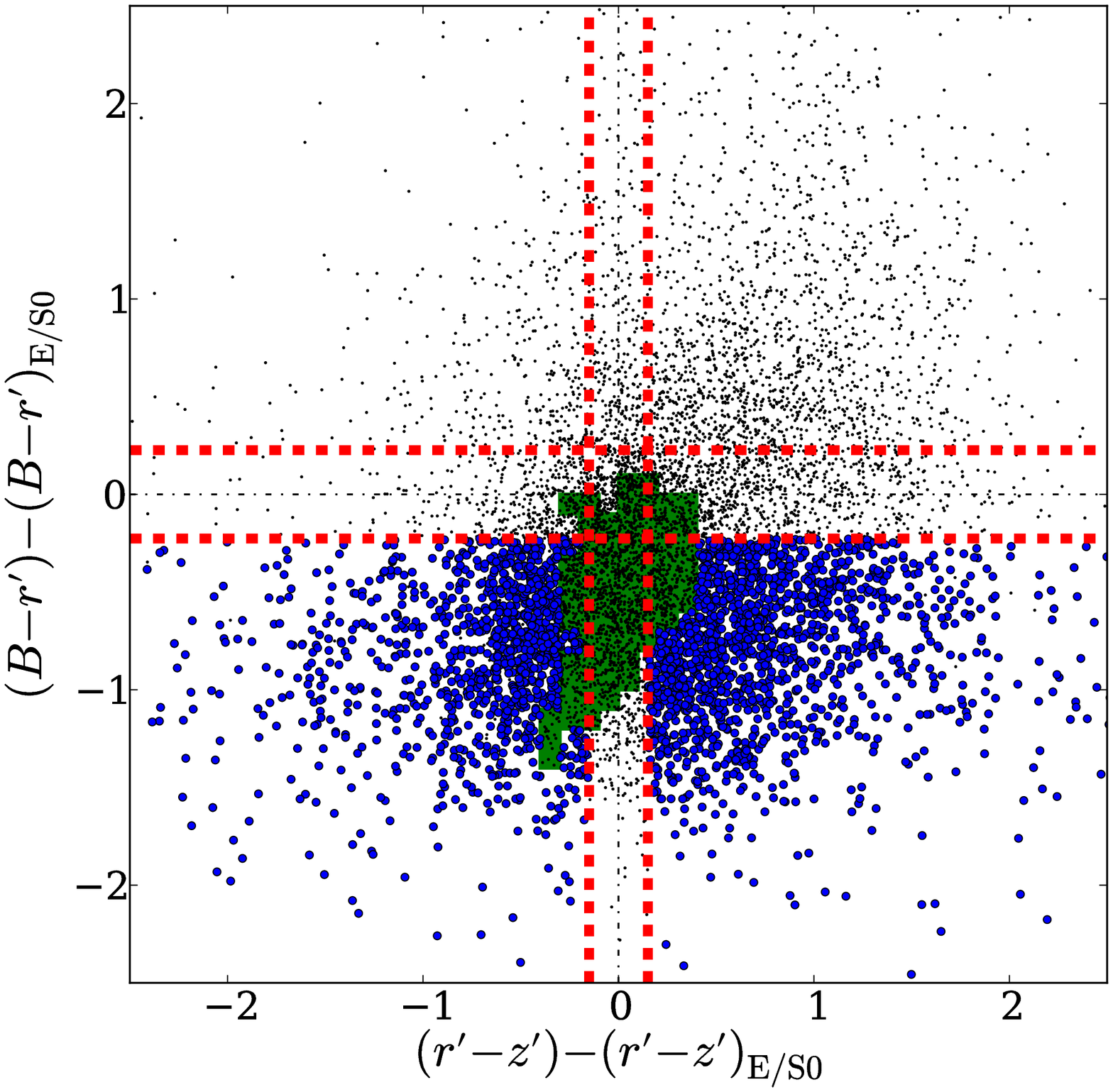}
\caption{Left :  $\langle D_{ls}/D_s\rangle$ distribution in the
 color-offset plane of $(r'-z')-(r'-z')_{\rm E/S0}$ and
 $(B-r')-(B-r')_{\rm E/S0}$. Right: black points are all galaxies for
 which shapes are measured. Blue points are background galaxies. 
Green region denotes the color area with $\langle
 D_{ls}/D_s\rangle<0.6$. Red dashed lines are the width of the red sequence.
}
\label{fig:selection}
\end{figure*}

A secure background selection is critically important to
measure accurately cluster and subcluster masses \citep[e.g.][]{Broadhurst05,Okabe13,Okabe15c}.
We use color information ($r'-z'$ and $B-r'$) for the selection 
\citep[e.g.][]{Medezinski10,Umetsu10}. 
The red-sequence of member E/S0 galaxies appeared in ($r'$ and $r'-z'$) and ($r'$ and
$B-r'$) planes is fitted with a linear function.
The mean lensing depth $\langle D_{ls}/D_s\rangle$ is
computed as a function of the color offsets from the red-sequence of
$(r'-z')-(r'-z')_{\rm E/S0}$ and $(B-r')-(B-r')_{\rm E/S0}$. 
Here, $D_{s}$ and $D_{ls}$ are the angular diameter distances to the sources and 
between the cluster (lens) and the sources, respectively.
The resulting $\langle D_{ls}/D_s\rangle$ map
is shown in the left panel of Figure \ref{fig:selection}. 
The grid box size is $0.1$ABmag. 
$\langle D_{ls}/D_s\rangle$ at small color-offsets presents 
low lensing depth, suggesting a finite contamination of member
galaxies. Following \cite{Medezinski10}, we first select galaxies of
which $(B-r')$ is bluer than those of clusters ($(B-r')-(B-r')_{\rm
E/S0}<-\sigma_{B-r'}$), 
where $\sigma_{B-r'}$ is the color width of the red sequence.
In a similar manner, galaxies of which colors are located in the width
of $\sigma_{r'-z'}$ are excluded in our source background catalog.
Since a combination of the lensing depth and the color information 
is a good estimator to 
monitor the color distribution for possible member galaxies,
we exclude color regions for our background sources by employing
 $\langle D_{ls}/D_s\rangle<0.6$, which is equivalent to $\langle z_s \rangle<0.52$.
The number density of background source galaxies is $2.6~{\rm arcmin^{-2}}$.
The mean source redshift is $\langle z_s\rangle=0.639$.

\section{Galaxy Distribution}\label{sec:gal}

\begin{figure*}
\includegraphics[width=0.97\hsize]{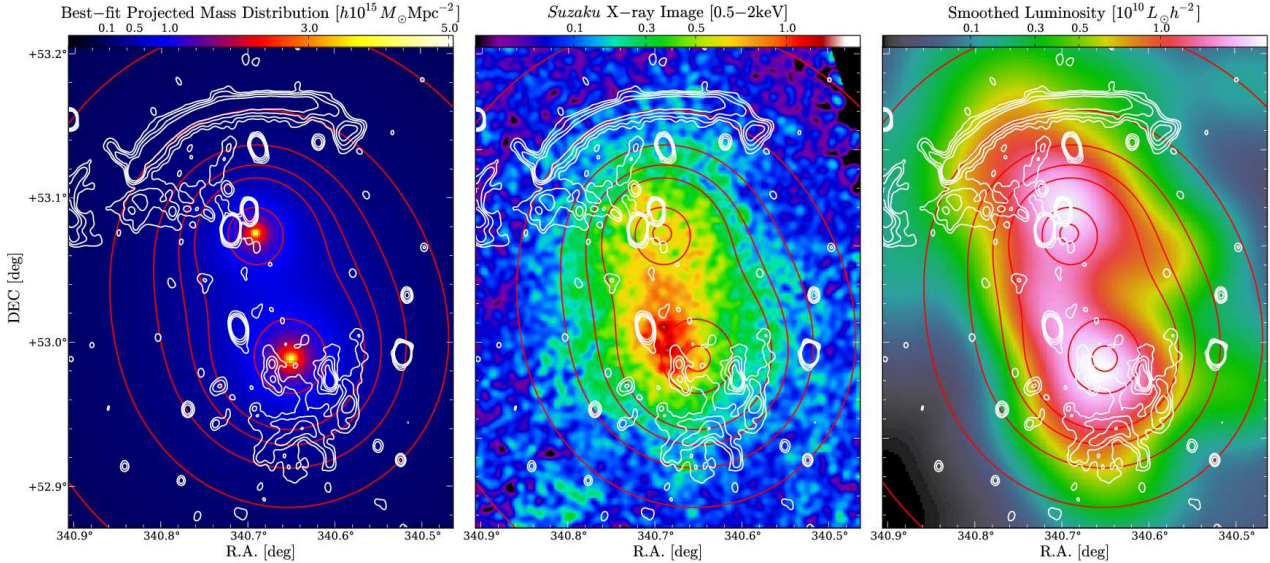}
\caption{Right: the best-fit mass distribution computed from a
 simultaneous fit of two-dimensional shear pattern (Sec
 \ref{sec:mass}). Contour (red) levels are in units of the convergence,
 $\kappa$, starting from $\kappa=0.1$ with steps of $\kappa=0.1$.
Overlaid contours (white) are the radio emission \citep{Stroe13} at
 $1.2$ GHz in units of signal-to-noise ratio $[4,8,16,32]\sigma_{\rm RMS}$.
Middle: X-ray image of {\it Suzaku} satellite in $0.5-2.0$keV \citep{Akamatsu13,Akamatsu15}.
Right : Luminosity map for bright member galaxies (Sec \ref{sec:gal})
 computed with Gaussian smoothing FWHM$=3\farcm3$.}
\label{fig:maps}
\end{figure*}

We make a map of member galaxies for visual purpose to depict the configuration of cluster merger.
We here define bright member galaxies with
$r'<21$ABmag, $|(B-r')-(B-r')_{\rm E/S0}|<\sigma_{B-r'}$ and
$|(r'-z')-(r'-z')_{\rm E/S0}|<\sigma_{r'-z'}$.
The luminosity in the $r'-$band is computed with 
K-correction assuming a single cluster redshift.
We make a luminosity map convolved with a 
Gaussian smoothing kernel $\propto \exp[-\theta^2/\theta_g^2]$,
where $\theta_g=2'$ corresponding to FWHM$=3\farcm3$.
The right panel of Figure \ref{fig:maps} clearly shows a bimodal
structure of member galaxies. 
The luminosity peaks are associated with the brightest cluster galaxies
for two halo components, which provides us with
 priors of the number of halos and their central positions.

\section{Mass Measurement} \label{sec:mass}

We simultaneously fit a two-dimensional shear pattern with multi-components
model, following previous papers \citep[see details;][]{Okabe11,Watanabe11,Merluzzi15}.
We pixelize the shear pattern into a regular grid of
$1'\times 1'$ without any spatial smoothing procedures.
We employ the NFW mass model predicted by numerical simulations
\citep{NFW96} as the mass model. 
The NFW mass density profile is specified by two parameters of the mass $M_\Delta$ and the halo
concentration $c_{\Delta}$, where $\Delta$ denotes $\Delta$ times 
the critical mass density, $\rho_{\rm cr}(z_l)$, at the cluster redshift.
We also treat centers ($\alpha_c,\delta_c$) of the two mass components as free parameters \citep[e.g.][]{Oguri10b}.
Based on a prior of member galaxy distribution (Section
\ref{sec:gal}; right panel of Figure \ref{fig:maps}), we consider
two-halo components in the northern and southern region. 
There are eight parameters in total $\mbox{\boldmath $p$}=(M_{\rm \Delta}^{\rm
S},c_{\rm \Delta}^{\rm S},M_{\rm \Delta}^{\rm N},c_{\rm \Delta}^{\rm N},
\alpha_c^{\rm S},\delta_c^{\rm S}, \alpha_c^{\rm N},\delta_c^{\rm N})$.
Here, the superscripts denote the southern halo (S) and the northern halo
(N), respectively.
We adopt the Markov Chain Monte Carlo method with standard
Metropolis-Hastings sampling. 
Since it is difficult to determine the concentration parameter of
each individual halo because of the small number of background
galaxies and the degeneracy between model parameters, we use a
Gaussian prior that follows the halo
concentration-mass ($c$--$M$) relation of \cite{Bhattacharya13}
derived from N-body cosmological simulations. The Bhattacharya et al.
(2013) $c$--$M$ relation is in agreement with recent stacked lensing
studies \citep{Okabe13,Okabe15c,Umetsu14,Umetsu15b}.
We use a flat prior for other parameters. 
The ranges of virial masses are restricted to $0<M_{\rm \Delta}<30\times10^{14}\hMsol$.
The centers are allowed within $\pm2\farcm$ around the candidates of
most luminous cluster galaxies.
The best-fit parameters are shown in Table \ref{tab:mass}.
The best-fit mass for the southern halo is twice higher than that for the
northern halo, though an equal-mass ratio cannot be ruled out within errors.
The offsets between lensing centers and most luminous galaxies on the sky plane
for the southern and the northern halos are $36^{+118}_{-36}\hkpc$ and $151_{-67}^{+233}\hkpc$, respectively. 
Thus, the offset is not significant for the southern main halo and at $\sim2\sigma$ level for the
northern subhalo.

\begin{figure}
\includegraphics[width=0.97\hsize]{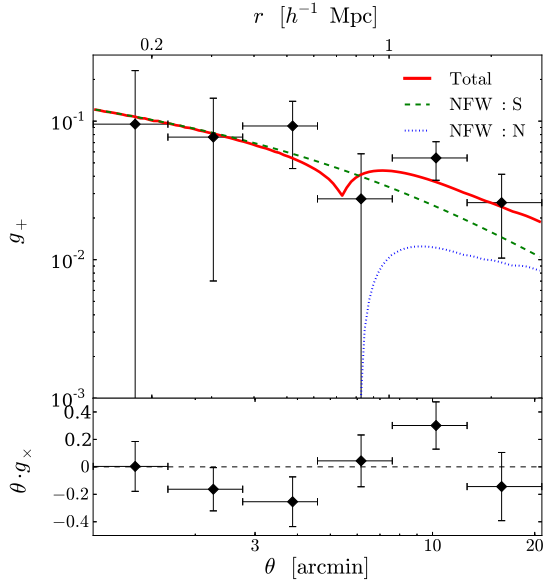}
\caption{Top panel : tangential shear profile as a function of the
 projected cluster-centric radius from the center of the southern main
 halo determined by two-dimensional shear fitting.
A bump in the profile signals is found at $r\simgt8'$. 
The profile is well described by two NFW components of the northern and
 southern halos. The red solid, green dashed
and blue dotted lines are the best-fitting NFW profile for the two halos, the southern halo and the northern halo, respectively. 
Bottom panel: the $45^\circ$ rotation components, $\theta\cdot g_\times$.}
\label{fig:g+}
\end{figure}

The tangential shear profiles of two mass components as a function of
the projected distance from the best-fit center of the southern halo are shown in Figure
\ref{fig:g+}. The center is set to be the best-fit value.
The lensing signal at $r\sim8\farcm$ is depressed by the northern mass halo.
The best-fit lensing profile for the southern halo (green dashed line)
monotonically decreases, while the best-fit lensing profile for the
northern halo (blue dotted line) rapidly increases up at
$r\sim8\farcm$ due to an off-centering effect \citep[e.g.][]{Yang06}.
The total mass model (red solid line) well describes the observed lensing profile.

\begin{table*}
\caption{Mass estimates for the southern and northern mass components
 ($10^{14}\hMsol$) and
 their centroids in the units of degree, respectively.} \label{tab:mass}
\begin{center}
\begin{tabular}{lcccccc}
\hline
  $\Delta$          & $M_{\rm \Delta}^{\rm S}$ 
                 & $c_{\rm \Delta}^{\rm S}$ 
                 & $M_{\rm \Delta}^{\rm N}$ 
                 & $c_{\rm \Delta}^{\rm N}$ 
                 & $(\alpha_c,\delta_c)^{\rm S}$
                 & $(\alpha_c,\delta_c)^{\rm N}$ \\
\hline
  vir         &  $12.44_{-6.58}^{+9.86}$
              &  $4.60_{-1.28}^{+1.96}$
              &  $6.74 _{-4.12}^{+7.64}$
              &  $5.34_{-1.61}^{+2.25}$
              &  $(340.651_{-0.019}^{+0.028},52.989_{-0.016}^{+0.015})$
              &  $(340.692_{-0.028}^{+0.020},53.076_{-0.014}^{+0.011})$ \\
$200$         &  $10.69_{-5.67}^{+9.82}$
              &  $3.44_{-0.98}^{+1.51}$
              &  $5.51_{-3.43}^{+6.39}$
              &  $4.11_{-1.26}^{+1.97}$
              &  $(340.651_{-0.019}^{+0.028},52.989_{+0.015}^{-0.016})$
              &  $(340.692_{-0.029}^{+0.020},53.075_{+0.012}^{-0.015})$\\
\hline 
\end{tabular}
\end{center} 
\end{table*}

\section{A comparison between Projected Distributions of Radio Relic, ICM, and Mass}

The acceleration process of relativistic electrons emitting giant relics
are believed to be associated with on-going merger, but the details remain unknown.
Since the dark matter controls the merger dynamics, 
a comparison of projected distributions between the dark matter, the gas
and the radio relics 
would provide us with an important information 
for understanding the non-thermal physics \citep[e.g.][]{Okabe10b}.
Since the number density of background galaxies is small, we compute
the projected mass distribution calculated from the best-fit parameters 
without any smoothing kernels, rather than weak-mass reconstruction
\citep{Okabe08} in order to avoid noisy feature. 
The best-fit projected mass distribution 
makes it clear to understand the configuration of cluster merger.
We also investigate the projected mass distribution by changing 
parameters with $1\sigma$ measurement errors, the overall feature is
not significantly changed.
The resultant maps are shown in the left panel of Figure \ref{fig:maps}.
The left panel is the best-fit mass distribution, overlaid with radio
contours \citep{Stroe13}. 
The more and less massive clusters are located at the south and the
north, respectively.
Both the northern and the southern relics are very similar to the
projected mass distribution. 
In particular, the curved shapes of two relics are tightly
correlated with the curvature of the iso-contours of the projected mass
distribution.
It indicates that the bimodal mass distribution with a different mass
ratio has some influence on the formation of asymmetric radio relics.

The middle panel compares the X-ray surface brightness
distribution with the mass distribution and the radio relics.
The X-ray distribution is elongated along the north-south
direction. 
The overall gas distribution is in an excellent agreement with the mass
distribution. 
Especially, the faint X-ray emission (green color) at $\sim5'$ from the merger axis
follows the curvature of dark matter distribution.
The X-ray peak is associated with the southern massive cluster, 
while no prominent gas core is found around the center of the northern less massive cluster. 
It suggests that the gas core associated with the northern subcluster is
disrupted by the merger \citep{Tormen04}.

The smoothed luminosity map for member galaxies (right panel;
Sec. \ref{sec:gal}) shows a clear bimodal spatial distribution 
which is similar to weak-lensing mass distribution. 
The luminosity around each central region is comparable to each other.
We compute the mass-to-light ratio within $<250h_{70}^{-1}$ with
central positions determined by two-dimensional shear pattern.
We obtain $M/L(<250h_{70}^{-1})=687\pm245h_{70}M_\odot/L_\odot$ for the
southern halo and $429\pm184h_{70}M_\odot/L_\odot$ for the northern halo,
respectively. Here, we use the projected mass for each halo component and the $r$-band
total luminosity corrected with a Schechter luminosity function 
in the central region \citep[$<500h_{70}^{-1}$;][]{Popesso05}.
The mass-to-light ratio is somewhat larger than those in 
other clusters \citep[e.g.][]{Okabe08,Medezinski10}. 
This could be caused by the fact that some member galaxies are missing due to a
contamination of a considerably number of stars in Galactic-plane position. 
Although the viral mass for the southern halo is about two times higher
than that for the northern halo, the mass-to-light ratios are comparable
because two masses within small radii are comparable due to a difference
of the concentration parameter (Table \ref{tab:mass}).

\section{Discussion}

The best-fit parameters obtained by two-dimensional shear fitting suggest that CIZAJ2242.8+5301 undergoes
the cluster merger with $\sim2:1$ mass ratio. The main cluster is the
southern halo and the subcluster is the northern halo, respectively.
The overall distribution of X-ray emitting gas 
is similar to the projected mass distribution (the middle panel in Figure \ref{fig:maps}). 
On the other hand, the gas core is associated with the main cluster, 
while no significant gas core around the northern subhalo is found.
It indicates that the gas core associated with the main cluster
survives under the collisional matter interaction and 
the gas core with the subhalo is disrupted \citep{Tormen04}. 
{\it Chandra} X-ray observation \citep{Ogrean14} has reported possible density jumps in the
surface brightness profile at inner radii between two relics. 
Visual inspection of their result suggests that those jumps follow the
overall mass distribution.
The {\it Chandra} X-ray image also shows that 
the X-ray emission from the northern central region (at $r\sim2'$ from
from the approximate northern subhalo center) is somewhat more extended
than that from the southern central region. It would point out that the
disrupted gas core in the subhalo is more extended.
Although the gas distribution does not always follow the dark
matter distribution in violent merging clusters \citep{Okabe08}, the
gas distribution is similar to the dark matter distribution in 
the cluster on a very early stage after the core passage.
The big giant radio relic is in the front of an advancing direction of
the northern subhalo. Asymmetric radio relics are very similar to
the projected mass distribution.

Since the dark matter subhalo is less disrupted by the cluster merger 
compared to the collisional gas core, 
the positions of dark matter halos provide us with the
fundamental information of the cluster merger.
The distance between weak-lensing determined centers of two halos is $\sim712\hkpc$. 
The center of mass can be estimated to be $(340.667{\rm
deg},53.0202{\rm deg})$ assuming a point mass. 
The distances between outer edge of the northern relic and the center of
mass, $d_{\rm NR-CM}$, and between outer edge of the southern relic and
the center of mass, $d_{\rm SR-CM}$ are $\sim9\farcm$ and $\sim5\farcm4$, respectively.
Interestingly, the distance ratio is
approximately equal to the ratio of the Mach numbers estimated by X-ray temperature, 
$d_{\rm NR-CM}:d_{\rm SR-CM} \sim {\mathcal M}_{\rm NR}:{\mathcal M}_{\rm SR}$ \citep{Akamatsu15}, where ${\mathcal M}_{\rm NR}=2.7^{+0.7}_{-0.4}$ and ${\mathcal
M}_{\rm SR}=1.7^{+0.4}_{-0.3}$. 
Furthermore, the projected distances between the dark matter halos and
the center of the mass,  $d_{\rm NM-CM}$ and $d_{\rm SM-CM}$, follow
this ratio $d_{\rm NM-CM}:d_{\rm SM-CM}\sim {\mathcal M}_{\rm NR}:{\mathcal M}_{\rm SR}$.
Since the distance is equivalent to the merger timescale multiplied by
the velocity, those positions relative to the center of mass 
are simply explained by the timescale after the core passage. 
In other words, the formation of radio relic is possibly triggered after the core passage. 
Our mass measurement could consistently explain the merger configuration
observed by X-ray, radio and weak-lensing techniques.

The particle acceleration mechanism of relativistic electrons is one of
the unsettled problems. 
A visual inspection of the sharp morphology of radio relics
\citep[e.g.][]{Brunetti14} and the spectral aging \citep{vanWeeren10} 
suggest that radio relics are connected with merger shocks.
Indeed, X-ray observations \citep{Akamatsu13,Akamatsu15} discovered the temperature
jump around the radio relics, though
the Mach numbers estimated by X-ray observation 
(${\mathcal M}_{\rm NR}=2.7\pm0.5$ and ${\mathcal M}_{\rm SR}=1.7\pm0.3$).
are somewhat lower than those (${\mathcal M_{\rm NR}}=4.58\pm1.09$ and ${\mathcal M_{\rm
SR}}=2.81\pm{0.19}$) by radio spectral index based on DSA model \citep{Stroe13}. 
\cite{Stroe14b} have revised ${\mathcal M_{\rm
NR}}=2.90^{+0.10}_{-0.13}$ by considering a convolution of the beam size of
radio telescopes.
Weak-lensing analysis has shown that the morphology of radio relics is
similar to the projected mass distribution.
It indicates that the shapes of radio relics on the cluster scale
radio relics well follow the global structure of the mass distribution.
The observational fact implies 
the connection between the gas physics and the dark matter distribution. 
Two scenarios of the formation of radio relic are proposed.
First, \cite{vanWeeren10} have suggested that a cluster merger shock 
accelerates electrons through some processes like the DSA \citep{Drury83}. 
Second, the fossil radio plasma seeded in
the past (e.g. radio lobes from active galactic nuclei) 
is re-accelerated by gas compression or shocks
\citep[e.g. ][]{Ensslin01,Hoeft04,Markevitch05}. \cite{Hoeft04} have shown by
numerical simulations that
the radio ghost is flared at curved merger shocks.
Although this study cannot constrain the formation process
of radio relics, it is worth mentioning that 
the second scenario should be discussed in the cluster community, 
because the first scenario conflicts with other observational evidences but the second does not.
Prominent detached shocks in A520 and the bullet cluster which have been
first discovered by X-ray observations are not associated with radio relics but with radio
halos \citep{Govoni04}. If all merger shocks injected relativistic
electrons, the radio relic should be observed in other famous merging clusters. 
If the formation of radio relic depends on the initial fossil radio
plasma, we do not face the above problem.
In order to understand the formation of radio relics, further
systematic multi-wavelength studies of X-ray, radio, optical and
weak-lensing analyses are vitally important.

We compare with previous numerical simulations
regarding cluster mergers. 
\cite{vanWeeren11} have conducted numerical simulations for the
cluster and shown that the northern massive cluster with the largest
X-ray core radius triggers the largest shock wave after the core passage. 
Their result conflicts with both X-ray and weak-lensing measurements, 
because the X-ray gas core is not in the north region and the
massive subhalo is located in the south region. 
Some discrepancy between the results of simulations and observations
would be caused by a choice of their initial conditions. For instance, 
they assumed that the gas temperature is almost isothermal, 
while recent studies \citep{Okabe14b} pointed out 
that the gas temperature universally drops off beyond $\sim0.5r_{200}$.
The intrinsic scatter of the gas density deviated from the universal
profile \citep{Okabe14b} might be also important. 
The observed configuration of the radio relics and X-ray surface
brightness distributions are realized by other numerical simulations \citep{Loken95, Roettiger99}.
A head-on merger with a mass ratio of 1:4 in less than 1Gyr after core
passage \citep{Loken95} have shown that 
the largest gas core and the prominent shock are 
found in the more and less massive halos, respectively.
\cite{Roettiger99} have conducted the magneto-hydrodynamics (MHD) and dark
matter simulation to study A3667 and shown 
that a cluster merger of a mass ratio of 1:5 reproduce a largest radio
relic associated with a subhalo and X-ray emission associated with a main halo. 
The mass ratios in two simulations are somewhat larger than that for the cluster.
Those simulations assumed the first order Fermi acceleration to compute the radio relics.
In order to precisely realize the
observed features by numerical simulations, further studies using
observational results would be important to understand the formation
of radio relics. Cosmological MHD simulations
\citep{Skillman11,Skillman13} would be also important to understand the
statistical properties of the radio relics.

In contrast to giant radio relics, no prominent radio halo is found in
 CIZAJ2242.8+5301 \citep{vanWeeren10,Stroe13} in the
central region even at lower frequency $152$MHz. 
That means that relativistic electrons are neither fully accelerated nor
 re-accelerated.
For example, dying relativistic electrons escape
from the backward side of the moving direction of radio relics,
but are not re-accelerated by downstream turbulence.
The fact leads to the condition 
\begin{eqnarray}
 t_{\rm acc} > {\rm min}\{t_{\rm cool},t_{\rm merger}\}, \label{eq:balance}
\end{eqnarray}
where $t_{\rm acc}$, $t_{\rm cool}$ and $t_{\rm merger}$ are the
acceleration timescale, the radiative cooling timescale and the merger
time scale related to the particle acceleration, respectively.
The Lorentz factor of relativistic electrons is $\gamma\sim
6.6\times10^{3}\left(\frac{\nu(1+z_l)}{153{\rm MHz}(1+0.192)}\right)^{1/2}\left(\frac{B}{1\mu{\rm G}}\right)^{-1/2}$.
Given the Lorentz factor,
the cooling time of relativistic electrons is computed by the
synchrotron and the inverse Compton losses, as follows,
\begin{eqnarray}
 t_{\rm cool}&\simeq&0.2
  \left(\frac{\gamma}{6\times10^{3}}\right)^{-1} \nonumber \\
&\times&\left[\left(\frac{B}{1\mu{\rm
					    G}}\right)^2+\left(\frac{B_{\rm
					    CMB}}{3.25(1+z)^2\mu {\rm G}}\right)^2\right]^{-1}\,{\rm Gyr},
\end{eqnarray}
where $B_{\rm CMB}^2/(8 \pi)$ is the energy density of the cosmic microwave
background (CMB).  
We define the merger timescale, $t_{\rm merger}$, as the time that 
the northern subhalo travels from the center-of-mass, 
\begin{eqnarray}
t_{\rm merger}\simeq0.2\left(\frac{d_{\rm
			NM-CM}}{440\hkpc}\right)\left(\frac{v}{2000{\rm  km~s^{-1}}}\right)^{-1}\,{\rm Gyr},
\end{eqnarray}
where we assume that the merging velocity $v=2000~{\rm  km~s^{-1}}$ is equivalent to the shock
velocity estimate by X-ray observation.
We assume stochastic acceleration due to gyroresonant interaction with Alfv\'en
waves \citep[e.g.][]{Ohno02,Fujita03,OSullivan09}, in the framework of
the second-order Fermi acceleration.
The acceleration timescale is computed by  
\begin{eqnarray}
 t_{\rm acc}\simeq \frac{K_g}{v_{\rm A}^2}, 
\end{eqnarray}
where $v_A$ is the Alfv\'en velocity and $K_g\simeq c r_{\rm L}/\xi_{\rm
res}$ 
is a spatial diffusion coefficient parallel to the mean magnetic field by a pitch angle scattering,
respectively. Here, $r_{\rm L}$ is the Larmor radius and $\xi_{\rm res}=(\delta B_{\rm res}/B)^2$ is 
a fractional energy of MHD turbulence at the
wavelength satisfying a resonance scattering condition. 
For simplicity, we ignore the wave damping effect. 
The balance between the cooling time/merger time and the acceleration time
constrains the fractional energy of the turbulence of 
\begin{eqnarray}
 \xi_{\rm res}^{\rm cool}&<& 8\times 10^{-7}  \left(\frac{B}{1\mu{\rm
G}}\right)^{-4}\times \nonumber \\
&&\left[\left(\frac{B}{1\mu{\rm					    G}}\right)^2+\left(\frac{B_{\rm
					    CMB}}{3.25(1+z)^2\mu {\rm
					    G}}\right)^2\right]\nonumber, 
\end{eqnarray}
and 
\begin{eqnarray}
 \xi_{\rm res}^{\rm merger}< 6\times 10^{-7}  \left(\frac{B}{1\mu{\rm
G}}\right)^{-7/2} \left(\frac{t_{\rm merger}}{0.2\,{\rm Gyr}}\right)^{-1}. 
\nonumber 
\end{eqnarray}
We here assume the electron density $n_e=10^{-3}\,[{\rm cm}^{-3}]$ and
the mean molecular weight $\mu=0.59$. 
The two constraints of $\xi_{\rm res}^{\rm merger}$ and $\xi_{\rm res}^{\rm cool}$
are comparable to each other. 
We note that the two equations above have different dependencies on the
magnetic field strength, B. These constraints change slightly depending
on the choice of the B-field strength. Although the exact limits would
depend on the spatial position, our observations provide a crude upper
limit on the fractional energy of the MHD turbulence as $\xi_{\rm res}<{\mathcal O}(10^{-6})$.
It indicates that the MHD turbulence is significantly suppressed in the central
region, at this merging stage of the cluster.
In the above discussion, we ignore the mechanism to generate the
Alfv\'en waves. 
If the Alfv\'en waves is connected by the cascade of the fluid
turbulence, the balance between the energy loss and gain ($t_{\rm
acc}=t_{\rm cool}$) does not change but 
$t_{\rm acc}=t_{\rm merger}$ is modified to $t_{\rm acc}+t_{\rm cas}=t_{\rm
merger}$, where $t_{\rm cas}$ is the time scale of the cascade. 
As a result, the upper limit on $\xi_{\rm res}^{{\rm merger}}$ becomes
larger than that on $\xi_{\rm res}^{{\rm cool}}$. 
We also note that the estimation of the MHD turbulence depends on the
precise mechanisms of particle acceleration on the basis of microplasma
physics, and thus further studies regarding other various mechanisms are 
essential for a deeper understanding of radio halos.
If future deep radio observations discover very faint diffuse
emission, the constrains on the MHD turbulence will be much improved.

We compare our weak-lensing masses with previous study
\citep{Jee15}. 
\cite{Jee15} performed a weak-lensing mass estimate for the
cluster using a color-magnitude selected galaxy sample that contains
cluster red-sequence galaxies fainter than 22 ABmag. However, it has
been well established by earlier work \citep[e.g.][]{Broadhurst05,Okabe13,Okabe15c}
that such a background selection can lead to systematic dilution of
the weak-lensing signal, resulting in a substantial underestimate of the
cluster mass especially at inner radii.
By fixing the concentration parameter according to the mean $c$--$M$
relation of \cite{Duffy08} that is based on the WMAP five-year cosmology, they estimated $M_{200}^{\rm
S}=6.9^{+2.7}_{-1.8}\times10^{14}\hMsol$ and $M_{200}^{\rm N}=7.7^{+2.6}_{-2.3}\times10^{14}\hMsol$ for the southern and northern halos, respectively. Their mass
estimates for the northern and southern components, as well as their
total mass estimate, are compatible with our results within large
errors inherent in noisy weak-lensing measurements with a small number
of background galaxies. On the other hand, \cite{Jee15} concluded
that the cluster is the result of an equal-mass merger, whereas our
best-fit model supports a $\sim 2:1$ ratio merger. In order to
understand the origin of this discrepancy, we re-performed
two-dimensional shear fitting assuming the single scaling concentration of Duffy
et al. (2008) as done by Jee et al. (2015).  We obtained $M_{200}^{\rm
S}=8.1_{-4.1}^{+5.9}\times10^{14}\hMsol$ and $M_{200}^{\rm N}=6.7_{-4.1}^{+7.3}\times10^{14}\hMsol$ for the southern and northern halos, respectively. The
resulting mass ratio is thus reduced to $\sim1.2$, in better
agreement with their results. Therefore, their conclusions 
may be sensitive to the particular choice of the mean concentration--mass relation.

\section{Summary}

We have conducted weak-lensing analysis using three-band imaging of
Subaru/Suprime-Cam. 
We have carefully selected background galaxy populations  free of
contamination by unlensed cluster galaxies, following the color-color
selection method developed by \citep{Medezinski10}.
 Two dimensional shear
fitting enabled us to measure masses for two halos and their positions \citep{Okabe11}. 
We have found that the southern halo is the main cluster and the northern
halo is the colliding subcluster of which mass is about half of the
main cluster mass. Spatial offsets between luminous galaxies and the
weak-lensing-determined centers are at maximum $2\sigma$ level.
The mass-to-light ratios for two halos within $250h_{70}^{-1}{\rm kpc}$ from
weak-lensing-determined centers are comparable because the concentration
parameter for the subcluster is higher than that for the main cluster.

The distance ratio between outer edges of the two relics and the center of mass
are comparable to those between two halo positions and the center of mass,
as well as between the Mach numbers determined by X-ray observations. 
The gas halo is associated with the southern main halo, 
while no well-defined gas core is found in the northern subhalo.
Our mass measurement could consistently explain the merger configuration
observed by X-ray, radio and weak-lensing techniques. 
The result implies a possibility that the radio relic is formed after the core passage. 

We have found that the curved shapes of radio relic coincides well with the projected mass distribution, 
indicating that radio relics on cluster-scale well follow the global structure of the
mass distribution. 

No prominent radio halo is found, which enabled us to put 
an upper limit of the fractional energy of MHD
 turbulence at resonant wavenumber of $(\delta
 B/B)^2<\mathcal{O}(10^{-6})$, by comparing the acceleration timescale
 of the resonant acceleration with the cooling time or the merger time after the
 core passage.

\section*{Acknowledgments}

We appreciate the anonymous referee for the helpful comments.
We are grateful to N. Kaiser for developing the IMCAT package and making it publicly available.
We thank van Weeren R.~J. for helpful discussion during his stay in
Hiroshima University.
This work was supported by ``World Premier International Research Center
Initiative (WPI Initiative)`` and the Funds for the Development of Human
Resources in Science and Technology under MEXT, Japan, and 
 Core Research for Energetic Universe in Hiroshima University (the MEXT
 program for promoting the enhancement of research universities, Japan).
N. Okabe is supported by a Grant-in-Aid from the Ministry of Education, Culture, 
Sports, Science, and Technology of Japan (26800097).
H. Akamatsu is supported by a Grant-in-Aid for Japan Society for the
Promotion of Science (JSPS) Fellows (26-606).
Y.-Y. Zhang acknowledges support by the German BMWI through the
  Verbundforschung under grant 50\,OR\,1304.
K. Umetsu acknowledges partial support from 
the Ministry of Science and Technology of Taiwan
(grant MOST 103-2112-M-001-030-MY3).

\bibliographystyle{apj}
\bibliography{my,ciza2242}

\end{document}